\documentclass[prd,twocolumn,amsmath,amssymb]{revtex4}
\usepackage{graphicx}%
\def\pp{\pi^+\pi^-}
\def\kpi{K^-\pi^+}
\def\kk{K^+K^-}
\def\gg{\gamma\gamma}
\def\ll{\Lambda \bar \Lambda}

\begin{document}
\title{\bf \boldmath Measurements of the observed cross sections for exclusive
light hadron production in $e^+e^-$ annihilation at $\sqrt{s}=$ 3.773 and
3.650 GeV}
\author{
\vspace{0.35cm}
{\large BES Collaboration}\\
\vspace{0.35cm}
M.~Ablikim$^{1}$,              J.~Z.~Bai$^{1}$,               Y.~Ban$^{12}$,
X.~Cai$^{1}$,                  H.~F.~Chen$^{16}$,
H.~S.~Chen$^{1}$,              H.~X.~Chen$^{1}$,              J.~C.~Chen$^{1}$,
Jin~Chen$^{1}$,                Y.~B.~Chen$^{1}$
% ...
\\
\vspace{0.5cm}
{\it
$^{1}$ Institute of High Energy Physics, Beijing 100049, People's Republic of China\\
% ...
$^{12}$ Peking University, Beijing 100871, People's Republic of China\\
% ...
$^{16}$ University of Science and Technology of China, Hefei 230026, People's Republic of China\\
% ...
}}
\email{mahl@mail.ihep.ac.cn (H. L. Ma)}

\begin{abstract}
By analyzing the data sets of 17.3 pb$^{-1}$ taken at $\sqrt{s}=3.773$ GeV and
6.5 pb$^{-1}$ taken at $\sqrt{s}=3.650$ GeV with the BESII detector at
the BEPC collider, we have measured the observed cross sections for
12 exclusive light hadron final states produced in $e^+e^-$ annihilation at the two
energy points. We have also set the upper limits on the observed cross sections
and the branching fractions for $\psi(3770)$ decay to these final states at
90\% C.L.
\end{abstract}

\maketitle

\oddsidemargin  -0.2cm
\evensidemargin -0.2cm

\section{Introduction}
As the lowest mass charmonium resonance above the $D\bar D$ threshold,
the $\psi(3770)$ is expected to decay almost entirely into pure
$D\bar D$ pairs. However, there has been a ``long-standing puzzle'' that the
$\psi(3770)$ is not saturated by $D\bar D$ decays \cite{rzhc}. Recently, CLEO
Collaboration measured the $e^+e^-\to\psi(3770)\to$ non-$D\bar D$
cross section to be $(-0.01\pm0.08^{+0.41}_{-0.30})$ nb \cite{besson}.
While, BES Collaboration measured the branching fraction for $\psi(3770)\to$
non$-D\bar D$ decay to be $(15\pm5)\%$ \cite{brdd1,brdd2,pdg07},
which implies that
the $\psi(3770)$ may substantially decay into charmless final states.
Meanwhile many measurements about exclusive
$\psi(3770) \to$ non-$D\bar D$ decays are reported 
by BES \cite{bai,ablikim,kskl,rhopi,crshads} and CLEO
\cite{adam,adams,coans,huang,cronin,briere} Collaborations.
However, these results can not explain the discrepancy between the observed cross
sections $\sigma_{D\bar D}^{\rm obs}$ and $\sigma_{\psi(3770)}^{\rm obs}$
for $D\bar D$ and $\psi(3770)$ production.
To understand the discrepancy, one may directly compare
the observed cross sections for more exclusive light hadron final states at
the center-of-mass energies at 3.773 GeV and below 3.660 GeV,
excluding the contributions from $J/\psi$ and $\psi(3686)$ due to ISR
(Initial State Radiation) returns as well as $D\bar D$ decays.
From these cross sections, one can also obtain
some valuable information to understand the mechanism of the
continuum light hadron production.

Following Ref. \cite{crshads}, we report measurements of the observed
cross sections for the exclusive light hadron final states of
$\kk2(\pp)$, $2(\kk)\pp$, $p\bar p2(\pp)$, $4(\pp)$, $\kk 2(\pp)\pi^0$,
$4(\pp)\pi^0$, $\rho^0\pp$, $\rho^0\kk$, $\rho^0 p\bar p$,
$K^{*0}\kpi+c.c.$, $\ll$ and $\ll\pp$ produced in $e^+e^-$ annihilation
at $\sqrt{s}=$ 3.773 and 3.650 GeV in this Letter.
The measurements are made by analyzing the data sets of 17.3 pb$^{-1}$
taken at $\sqrt{s}=$ 3.773 GeV and 6.5 pb$^{-1}$ taken at
$\sqrt{s}=$ 3.650 GeV with the BESII detector at the BEPC collider.
For convenience, we call these two data sets
to be the $\psi(3770)$ resonance data and the
continuum data in the Letter, respectively.
 
\section{BESII detector}
The BESII is a conventional cylindrical magnetic detector that is
described in detail in Refs. \cite {bes,bes2}. A 12-layer Vertex Chamber (VC)
surrounding a beryllium beam pipe provides input to event trigger,
as well as coordinate information. A forty-layer main drift chamber
(MDC) located just outside the VC yields precise measurements of charged
particle trajectories with a solid angle coverage of $85\%$ of 4$\pi$;
it also provides ionization energy loss ($dE/dx$) measurements for
particle identification. Momentum resolution of $1.7\%\sqrt{1+p^2}$
($p$ in GeV/$c$) and $dE/dx$ resolution of $8.5\%$ for Bhabha scattering
electrons are obtained for the data taken at $\sqrt{s}=3.773$ GeV. An
array of 48 scintillation counters surrounding the MDC measures time
of flight (TOF) of charged particles with a resolution of about 180
ps for electrons. Outside the TOF counters, a 12 radiation length,
lead-gas barrel shower counter (BSC), operating in limited streamer
mode, measures the energies of electrons and photons over $80\%$ of
the total solid angle with an energy resolution of $\sigma_E/E=0.22
/\sqrt{E}$ ($E$ in GeV) and spatial resolutions of $\sigma_{\phi}=7.9$
mrad and $\sigma_Z=2.3$ cm for electrons. A solenoidal magnet outside
the BSC provides a 0.4 T magnetic field in the central tracking region
of the detector. Three double-layer muon counters instrument the magnet
flux return and serve to identify muons with momentum greater than 500
MeV/c. They cover $68\%$ of the total solid angle.

\section{Event Selection}
\label{evtsel}
We reconstruct $e^+e^-\to$ exclusive light hadrons mentioned above
by selecting the final states $m(\pp)n(\kk)i(p\bar p)j(\gg)$
$(m=1,2,4;\hspace{0.1cm}n=0,1,2;\hspace{0.1cm}i=0,1\hspace{0.1cm}{\rm and}
\hspace{0.1cm}j=0,1)$ from the data.
For the processes containing $\pi^0$, $\rho^0$, $K^{*0}$ and
$\Lambda$ mesons in the final states, we reconstruct these intermediate
resonances by the decays $\pi^0\to\gg$, $\rho^0\to\pp$, $K^{*0}\to K^+\pi^-$
and $\Lambda\to p\pi^-$, respectively. For the decays $K^{*0}\to K^+\pi^-$
and $\Lambda\to p\pi^-$, charge conjugation is implied throughout the
Letter.

To select good candidate events, we require that at least four charged
tracks be well reconstructed in the MDC with good helix fits, and each of
them be within $|\rm{cos\theta}|<0.85$, where $\theta$ is the polar angle.
All charged tracks, save those from the $\Lambda$ decays, are required to
originate from the interaction region $V_{xy}<2.0$ cm and $|V_z|<20.0$ cm,
while those from the $\Lambda$ decays are required to originate from the
interaction region $V_{xy}<8.0$ cm and $|V_z|<20.0$ cm, where $V_{xy}$ and
$|V_z|$ are the closest approaches of the charged track in the $xy$-plane
and $z$ direction, respectively.

The charged particles are identified with the $dE/dx$ and TOF measurements,
with which the combined confidence levels ($CL_{\pi}$, $CL_K$ and $CL_p$)
for pion, kaon and proton hypotheses are calculated. The pion, kaon and
proton candidates are required to satisfy $CL_{\pi}>0.001$, $CL_K>CL_{\pi}$
and $\frac{CL_p}{CL_{\pi}+CL_K+CL_p}>0.6$, respectively.

The good photon candidates are selected with the BSC measurements by the
following criteria:
the energy of the photon deposited in the BSC is greater than 50 MeV,
the electromagnetic shower starts in the first 5 readout layers,
the angle between the photon and the nearest charged track is
greater than $22^\circ$ \cite{besdphy},
the opening angle between the cluster development direction and
the photon emission direction is less than $37^\circ$ \cite{besdphy}.

In each event, several different charged track and/or neutral particle combinations
may satisfy the above selection criteria for each exclusive light hadron
final state. Each combination is imposed on an energy-momentum conservation
kinematic fit. Only those combinations with a kinematic fit probability
greater than $1\%$ are accepted. If more than one combination satisfies the
selection criteria in an event, only the combination with the largest fit
probability is retained.

For the $2(\pp)$ final state, we veto the background events of
$(\gamma)J/\psi\pp$ with $J/\psi \to \mu^+\mu^-$ by requiring that
the invariant mass of any $\pp$ combination be less than 3.00
GeV/$c^2$, where $\mu^+\mu^-$ pairs may be misidentified as $\pp$ pairs.

For the $\kk\pp$, or $\kk2(\pp)[\pi^0]$ final state,
we exclude the events from $D\bar D$ decays by rejecting those in which
the $D$ and $\bar D$ mesons can be reconstructed in the decay modes of
$D^0\to \kpi$ and $\bar D^0\to K^+\pi^-$,
or $D^0 \to \kpi [\pi^0]$ and $\bar D^0 \to K^+\pi^-\pi^-\pi^+$,
or $D^+ \to \kpi\pi^+[\pi^0]$ and $D^-\to K^+\pi^-\pi^-$,
here charge conjugation is implied \cite{besdcrs}.
The contribution from other $D$ meson decay modes is accounted by using
Monte Carlo simulation, as discussed in Section \ref{backsub}.

\section{Data Analysis}
To study the final states $\kk2(\pp)$, $2(\kk)\pp$, $p\bar p2(\pp)$ and
$4(\pp)$, we define a kinematic quantity of the scaled energy as
$E_{\rm msr}/E_{\rm cm}$ \cite{crshads}. Here, $E_{\rm msr}$ is the total
originally measured energy of all the observed particles, and $E_{\rm cm}$
is the nominal center-of-mass energy. To look for possible intermediate
resonances, we examine the invariant mass spectra of the $\gg$, $\pp$,
$K^{\pm}\pi^{\mp}$ and $p\pi^-/\bar p\pi^+$ combinations from the
selected $\kk2(\pp)\gg$, $4(\pp)\gg$, $2(\pp)$, $\kk\pp$, $p\bar p\pp$
and $p\bar p2(\pp)$ events. In the Letter, they are denoted by
$M_{\gg}$, $M_{\pp}$, $M_{K^{\pm}\pi^{\mp}}$ and $M_{p \pi^-/\bar p\pi^+}$,
respectively.

\subsection{Candidates for $e^+e^- \to \kk2(\pp)$, $e^+e^- \to 2(\kk)\pp$,
$e^+e^- \to p\bar p2(\pp)$ and $e^+e^- \to 4(\pp)$}
\begin{figure}[htbp]
\includegraphics*[width=8.0cm]
{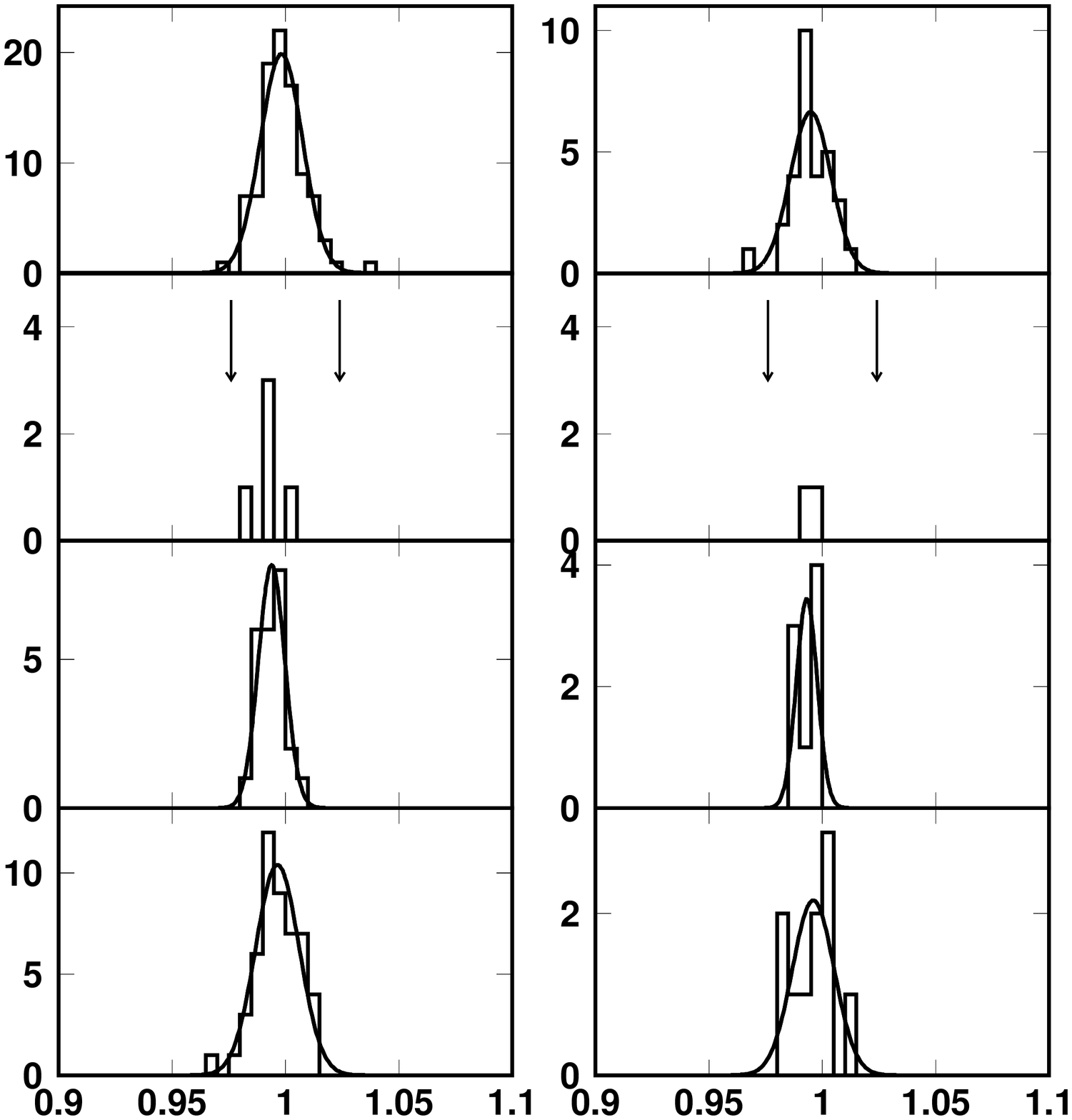}
\put(-125,0){\bf $E_{\rm msr}/E_{\rm cm}$}
\put(-235,88){\rotatebox{90}{\bf Events/(0.005)}}
\put(-135.5,207){\bf (a)}
\put(-135.5,158){\bf (b)}
\put(-135.5,109){\bf (c)}
\put(-135.5,60){\bf (d)}
\put(-38,207){\bf (a')}
\put(-38,158){\bf (b')}
\put(-38,109){\bf (c')}
\put(-38,60){\bf (d')}
\caption{
The $E_{\rm msr}/E_{\rm cm}$ distributions of the candidates for
(a) $e^+e^- \to \kk 2(\pp)$,
(b) $e^+e^- \to 2(\kk)\pp$,
(c) $e^+e^- \to p\bar p2(\pp)$ and
(d) $e^+e^- \to 4(\pp)$
selected from the $\psi(3770)$ resonance data (left) and the continuum data
(right), where the pairs of arrows show the signal regions.}
\label{fig:rat}
\end{figure}

Figure \ref{fig:rat} shows the resulting $E_{\rm msr}/E_{\rm cm}$
distributions, with each peak centered around unity correctly,
from the candidates for $e^+e^-\to \kk2(\pp)$,
$e^+e^- \to 2(\kk)\pp$, $e^+e^- \to p\bar p2(\pp)$ and
$e^+e^- \to 4(\pp)$. Fitting to the spectrum in each figure with a
Gaussian function for the signal and a flat background yields the number
$N^{\rm obs}$ of the signal events for each final state observed from the
$\psi(3770)$ resonance data and the continuum data. However, there are only a
few events in Figs. \ref{fig:rat}(b) and \ref{fig:rat}(b'). 
The mass window of $\pm 3\sigma_{E_{\rm msr}/E_{\rm cm}}$ around the
nominal $E_{\rm msr}/E_{\rm cm}$ is taken as the
$2(\kk)\pp$ signal region, where $\sigma_{E_{\rm msr}/E_{\rm cm}}$
is the resolution of the $E_{\rm msr}/E_{\rm cm}$ distribution determined by
Monte Carlo simulation. The pairs of
arrows in the figures show the $2(\kk)\pp$ signal regions.
In the signal regions, there are 5 and 2 signal
events for the $2(\kk)\pp$ final state observed from the $\psi(3770)$
resonance data and the continuum data, respectively.

\subsection{Candidates for $e^+e^- \to \kk2(\pp)\pi^0$ and
$e^+e^- \to 4(\pp)\pi^0$}
\begin{figure}[htbp]
\begin{center}
\includegraphics*[width=8.0cm]
{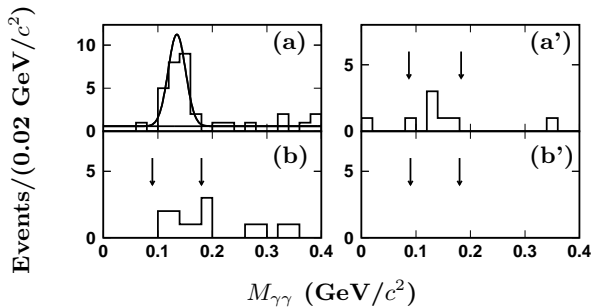}
\put(-145,0){\bf $M_{\gg}$ (GeV/$c^2$)}
\put(-235,10){\rotatebox{90}{\bf Events/(0.02 GeV/$c^2$)}}
\put(-135.5,95){\bf (a)}
\put(-135.5,52){\bf (b)}
\put(-38,95){\bf (a')}
\put(-38,52){\bf (b')}
\caption{
The distributions of the invariant masses of the $\gg$ combinations
from the candidates for
(a) $e^+e^- \to \kk 2(\pp)\pi^0$ and
(b) $e^+e^- \to 4(\pp)\pi^0$
selected from the $\psi(3770)$ resonance data (left) and the continuum data (right),
where the pairs of arrows indicate the $\pi^0$ signal regions.}
\label{fig:pi0}
\end{center}
\end{figure}

Figure \ref{fig:pi0} shows the distributions of the invariant masses of the
$\gg$ combinations from the candidates for $e^+e^- \to
\kk 2(\pp)\pi^0$ and $e^+e^- \to 4(\pp)\pi^0$.
In Fig. \ref{fig:pi0}(a), a $\pi^0$ signal is clearly observed.
Fitting to the invariant mass spectrum with a Gaussian function for the
$\pi^0$ signal and a flat background yields $21.3\pm5.0$ signal events for
$e^+e^- \to \kk2(\pp)\pi^0$ observed from the $\psi(3770)$ resonance data.
In the other figures, only a few events are observed, the mass window of
$\pm 3\sigma_{M_{\gg}}$ around the $\pi^0$ nominal mass is taken as the
$\pi^0$ signal region, where $\sigma_{M_{\gg}}$ is the $\pi^0$ mass
resolution determined by Monte Carlo simulation. In Fig. \ref{fig:pi0}(a'),
there are 6(2) events observed in the signal region(the outside of the
signal region). By assuming that the distribution of background is flat,
$0.6\pm0.4$ background event in the signal region is estimated. After
subtracting the background, we obtain $5.4\pm2.5$ signal events for
$e^+e^- \to \kk2(\pp)\pi^0$ observed from the continuum data. In Figs.
\ref{fig:pi0}(b) and \ref{fig:pi0}(b'), counting the events with $M_{\gg}$
within the $\pi^0$ signal region, we obtain the numbers $N^{\rm obs}$ of
the candidates for $e^+e^- \to 4(\pp)\pi^0$ from the $\psi(3770)$ resonance
data and the continuum data, respectively.

\subsection{Candidates for $e^+e^- \to \rho^0\pp$, $e^+e^- \to\rho^0\kk$
and $e^+e^- \to\rho^0 p\bar p$}
\begin{figure}[htbp]
\begin{center}
\includegraphics*[width=8.0cm]
{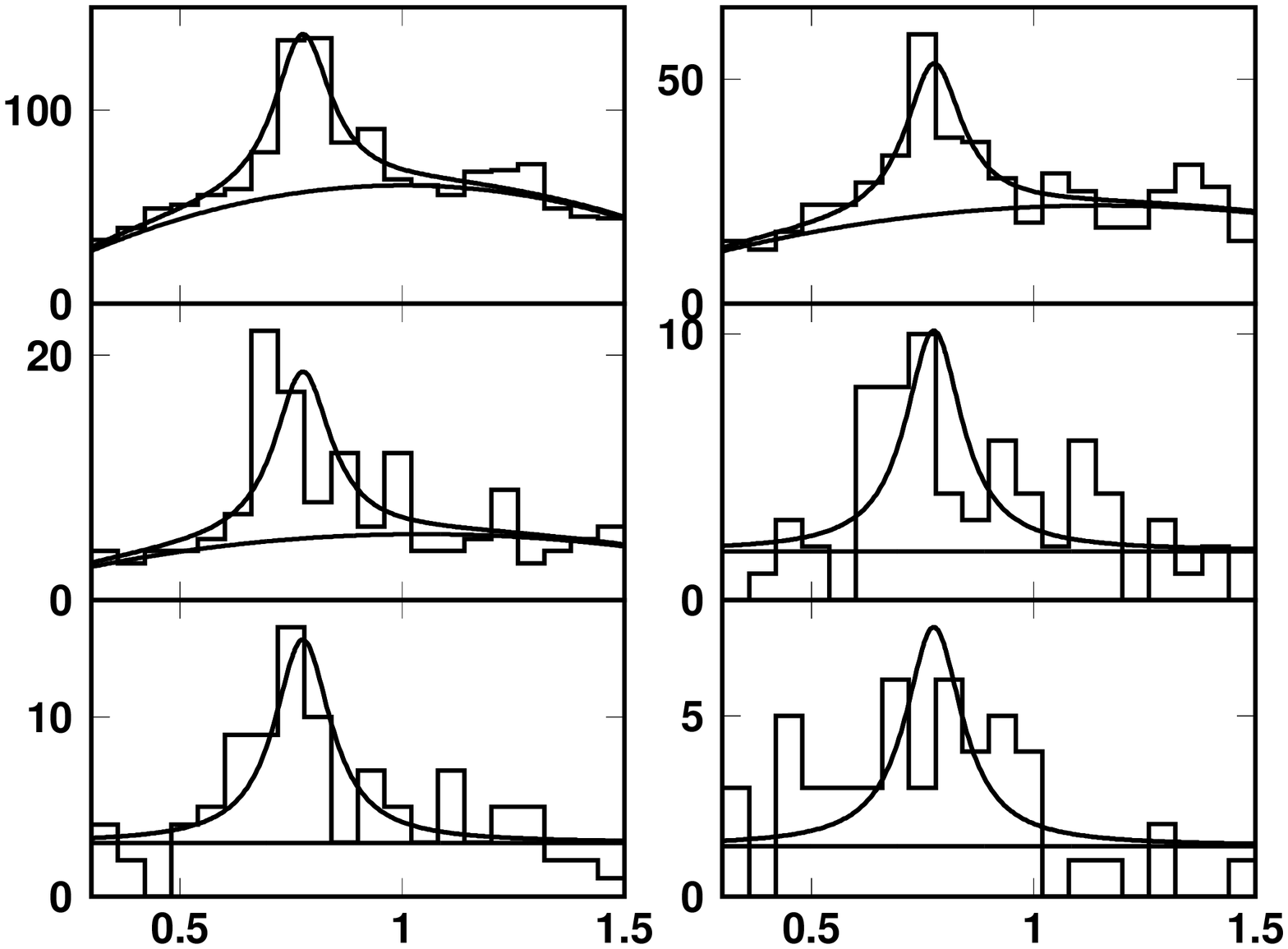}
\put(-145,0){\bf $M_{\pp}$ (GeV/$c^2$)}
\put(-235,45){\rotatebox{90}{\bf Events/(0.06 GeV/$c^2$)}}
\put(-135.5,150){\bf (a)}
\put(-135.5,104){\bf (b)}
\put(-135.5,58){\bf (c)}
\put(-38,150){\bf (a')}
\put(-38,104){\bf (b')}
\put(-38,58){\bf (c')}
\caption{
The distributions of the invariant masses of the $\pp$ combinations
from the selected
(a) $2(\pp)$,
(b) $\kk\pp$ and
(c) $p\bar p\pp$ events
from the $\psi(3770)$ resonance data (left) and the continuum data (right).}
\label{fig:xmpipi}
\end{center}
\end{figure}

To investigate the processes $e^+e^- \to \rho^0\pp$, $e^+e^- \to
\rho^0\kk$ and $e^+e^- \to\rho^0 p\bar p$, we analyze the
invariant mass spectra of the $\pp$ combinations from the selected
$2(\pp)$, $\pp\kk$ and $\pp p\bar p$ events, as shown in Fig.
\ref{fig:xmpipi}. The $\rho^0$ production can be observed in each
figure. Fitting to these invariant mass spectra with a Breit-Wigner
function convoluted with a Gaussian resolution function for the $\rho^0$
signal and a polynomial for the background, we obtain
the numbers $N^{\rm obs}$ of the signal events for $e^+e^- \to\rho^0\pp$,
$e^+e^- \to\rho^0\kk$ and $e^+e^- \to\rho^0 p\bar p$ observed from the
$\psi(3770)$ resonance data and the continuum data. In the fit, the
mass and width of $\rho^0$ are fixed to the world averaged values from
PDG \cite{pdg}.

\subsection{Candidates for $e^+e^- \to K^{*0}\kpi+c.c.$}
\begin{figure}[htbp]
\begin{center}
\includegraphics*[width=8.0cm]
{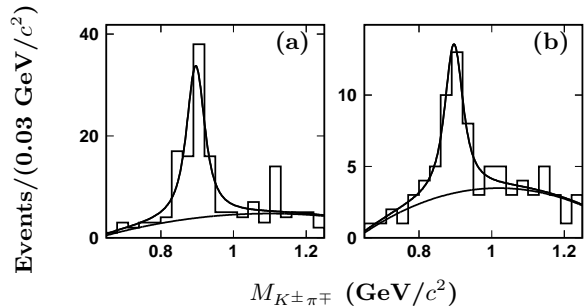}
\put(-145,0){\bf $M_{K^{\pm}\pi^{\mp}}$ (GeV/$c^2$)}
\put(-235,10){\rotatebox{90}{\bf Events/(0.03 GeV/$c^2$)}}
\put(-135.5,95){\bf (a)}
\put(-38,95){\bf (b)}
\caption{
The distributions of the invariant masses of the $K^{\pm}\pi^{\mp}$
combinations from the selected $\kk\pp$ events
(a) from the $\psi(3770)$ resonance data and (b) from the continuum data.}
\label{fig:xmkpi}
\end{center}
\end{figure}

To study the final state $e^+e^- \to K^{*0}\kpi+c.c.$,
we analyze the invariant mass spectra of the $K^{\pm}\pi^{\mp}$ combinations
from the selected $\kk\pp$ events, as shown in Fig. \ref{fig:xmkpi}.
In each figure, the $K^{*0}$ signal is clearly observed. Fitting to these
invariant mass spectra with a Breit-Wigner
function convoluted with a Gaussian resolution function for
the $K^{*0}$ signal and a second order polynomial for the background, we obtain
the numbers $N^{\rm obs}$ of the signal events for
$e^+e^-\to K^{*0}\kpi+c.c.$ observed from the $\psi(3770)$ resonance data
and the continuum data. In the fit, the mass and width of $K^{*0}$ are fixed
to the world averaged values from PDG \cite{pdg}.

\subsection{Candidates for $e^+e^-\to \ll$ and $e^+e^-\to \ll\pp$}
\begin{figure}[htbp]
\begin{center}
\includegraphics*[width=8.0cm]
{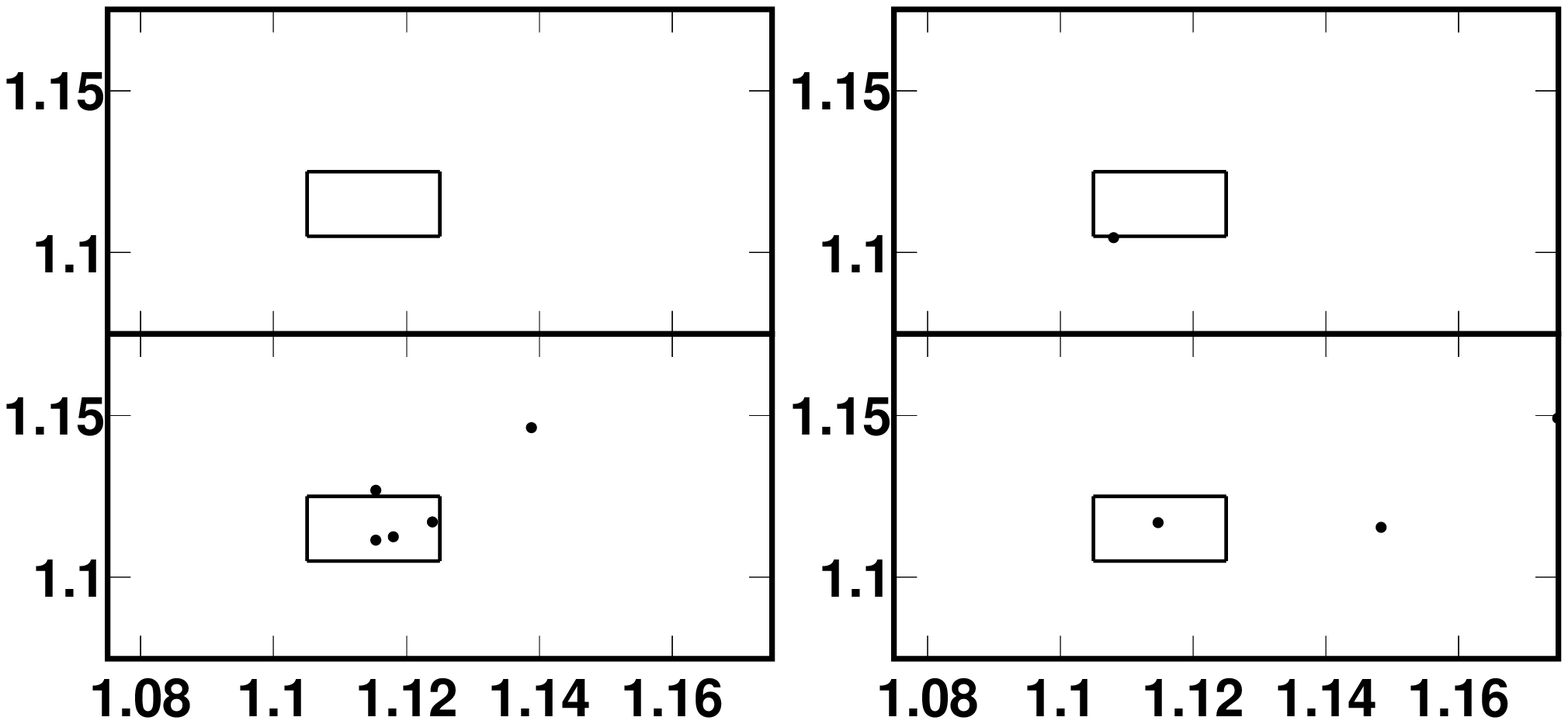}
\put(-145,0){\bf $M_{p \pi^-}$ (GeV/$c^2$)}
\put(-235,25){\rotatebox{90}{\bf $M_{\bar p \pi^+}$ (GeV/$c^2$)}}
\put(-135.5,95){\bf (a)}
\put(-135.5,52){\bf (b)}
\put(-38,95){\bf (a')}
\put(-38,52){\bf (b')}
\caption{
The scatter plots of $M_{p \pi^-}$ versus $M_{\bar p \pi^+}$ from
the selected 
(a) $p\bar p \pp$ and
(b) $p\bar p 2(\pp)$ events
from the $\psi(3770)$ resonance data (left) and the continuum data
(right), where the rectangle regions show the $\Lambda$ and $\bar \Lambda$
signal regions.}
\label{fig:xmppi_xmppi}
\end{center}
\end{figure}

We also inspect the scatter plots of $M_{p \pi^-}$ versus
$M_{\bar p \pi^+}$ from the selected $p\bar p\pp$ and $p\bar p2(\pp)$
events for studying $e^+e^-\to \ll$ and $e^+e^-\to \ll\pp$. They are shown
in Fig. \ref{fig:xmppi_xmppi}. In each figure, the mass window of 
$\pm 3\sigma_{M_{p \pi^-/\bar p\pi^+}}$ (10 MeV/$c^2$) around the
$\Lambda/\bar \Lambda$ nominal mass is taken as the $\Lambda/\bar \Lambda$
signal region.
Here, $\sigma_{M_{p \pi^-/\bar p\pi^+}}$ is the $\Lambda/\bar \Lambda$
mass resoluation determined by Monte Carlo simulation. In the signal
regions, there is 0(3) and 0(1) signal event for the $\ll$ ($\ll\pp$)
final state observed from the $\psi(3770)$ resonance data and the
continuum data, respectively.

\section{Background Subtraction}
\label{backsub}

For the selected candidate events, there are still the contributions from
$J/\psi$ and $\psi(3686)$ decays due to ISR returns,
the contributions from
the other final states due to misidentifying a $\pi$ as a $K$ or reverse,
and the contributions from $D\bar D$ decays. The number $N_{\rm b}$ of
these contributions should be subtracted from the number $N^{\rm obs}$ of
the candidate events for $e^+e^- \to f$ ($f$ represents exclusive light
hadron final state). These are estimated by Monte Carlo simulation,
which has been described in detail in Ref. \cite{crshads}.
Monte Carlo study shows that the other contributions
from the decays $\psi(3770) \to J/\psi\pp$, $\psi(3770) \to J/\psi\pi^0\pi^0$,
$\psi(3770) \to J/\psi\pi^0$, $\psi(3770) \to\gamma \chi_{cJ}\hspace{0.1cm}
(J=0,1,2)$ can be neglected \cite{crshads}.
For the $\kk\pp$ and $\kk2(\pp)[\pi^0]$ final states, even though we have
removed the main contributions from $D\bar D$ decays in the previous event
selection (see section \ref{evtsel}), there are still some events from
$D\bar D$ decays satisfying the selection criteria for the light hadron
final states. The number of these events should also be further removed
based on Monte Carlo simulation.

After the background subtraction, we obtain the net numbers of the signal
events, $N^{\rm net}$, for some final states. For the other
final states, only a few events are observed from the data. We set the upper
limits $N^{\rm up}$ on these numbers of the signal events by using the
Feldman-Cousins method \cite{felman} in the absence of background at 90\%
confidence level (C.L.). The numbers of $N^{\rm obs}$, $N^{\rm b}$ and
$N^{\rm net}$ (or $N^{\rm up}$) are summarized in Tabs. \ref{tab:crs3773}
and \ref{tab:crs3650}.

\section{Results}
\subsection{Monte Carlo efficiency}
The detection efficiency $\epsilon$ for $e^+e^-\to$ exclusive light hadrons
is estimated by Monte Carlo simulation for the BESII detector \cite{bessim}.
The Monte Carlo events are generated by using a phase space generator
including initial state radiation and vacuum polarization
corrections \cite{isr} with $1/s$ cross section energy dependence.
Final state radiation \cite{fsr} decreases the detection efficiency not
more than 0.5\%. The generator was used to determine detection
efficiency for $e^+e^-\to$ exclusive light hadrons in our previous work
\cite{crshads}. Detailed analysis gives the detection efficiency for each
final state at $\sqrt{s}=$ 3.773 and 3.650 GeV, as shown in the fifth
columns of Tabs. \ref{tab:crs3773} and \ref{tab:crs3650}. For the modes
containing intermediate resonances in the final states, the detection
efficiencies listed in the tables do not include the branching fractions
for the decays of the intermediate resonances.

\subsection{Systematic error}
\label{sys}
The systematic error in the measurement of the observed cross section
for the exclusive light hadron production arises mainly from the
uncertainties in
the integrated luminosity of the data set ($\sim$2.1\% \cite{brdd1,brdd2}),
the photon selection ($\sim$2.0\% per photon),
the tracking efficiency ($\sim$2.0\% per track),
the particle identification ($\sim$0.5\% per pion or kaon, $\sim$2.0\% per proton),
the kinematic fit ($\sim$1.5\%),
the Monte Carlo statistics ($\sim$(0.7$\sim$5.0)\%),
the branching fractions quoted from PDG \cite{pdg}
($\sim$0.03\% for $\mathcal B(\pi^0\to\gg)$ and
$\sim$0.78\% for $\mathcal B(\Lambda \to p\pi^-)$),
the background subtraction ($\sim$(0.0$\sim$3.5)\%),
the fit to mass spectrum ($\sim$(1.9$\sim$13.2)\%), and
the Monte Carlo modeling ($\sim$6.0\% \cite{crshads}).
The total systematic error $\Delta_{\rm sys}$ for each final state is
obtained by adding these uncertainties in quadrature.
They are summarized in the sixth columns of Tabs. \ref{tab:crs3773}
and \ref{tab:crs3650}.

\subsection{Observed cross sections for some final states}
The observed cross section for $e^+e^- \to f$ is obtained by dividing the
number $N^{\rm net}$ of the signal events by the integrated luminosity
$\mathcal{L}$ of the data set and the detection efficiency $\epsilon$
(and the branching fraction ${\mathcal B}_i$ for the possible
intermediate resonance decay $i$ containing in some final states),
\begin{equation}
\sigma_{e^+e^-\to f} = \frac{N^{\rm net}} {\mathcal{L}
\times \epsilon
\left (\times \prod_i^n {\mathcal B}_i \right )
},
\label{eq:crs}
\end{equation}
where the $n$ is the number of the intermediate resonances in the final
state, the ${\mathcal B}_i$ denotes
${\mathcal B}(\pi^0\to\gg)$,
${\mathcal B}(\rho^0\to\pp)$,
${\mathcal B}(K^{*0}\to K^+\pi^-)$ or
${\mathcal B}(\Lambda\to p\pi^-)$
\cite{pdg}.
Inserting the numbers of $N^{\rm net}$, $\mathcal{L}$, $\epsilon$
(and ${\mathcal B}_i$) in Eq. (\ref{eq:crs}), we obtain
$\sigma_{e^+e^-\to f}$ for some final states at $\sqrt{s}=$ 3.773 and 3.650
GeV, respectively. They are shown in the last columns of Tabs. \ref{tab:crs3773}
and \ref{tab:crs3650}, where the first error is statistical and the second systematic.

\subsection{Upper limits on observed cross sections for the other final states}
For the other final states, only a few events are observed from the data.
The upper limit on the observed cross section for $e^+e^- \to f$ is set by
\begin{eqnarray}
\sigma^{\rm up}_{e^+e^-\to f}=
\frac{N^{\rm up}} {\mathcal{L} \times
\epsilon
\times (1-\Delta_{\rm sys})
\left (\times \prod_i^n {\mathcal B}_i \right )},
\label{eq:crsup}
\end{eqnarray}
where $N^{\rm up}$ is the upper limit on the number of the signal events
for the final state and $\Delta_{\rm sys}$ is the systematic error in the
measurement of the observed cross section.
Inserting the numbers of $N^{\rm up}$, $\mathcal{L}$, $\epsilon$,
$\Delta_{\rm sys}$ (and ${\mathcal B}_i$)
in Eq. (\ref{eq:crsup}), we obtain $\sigma^{\rm up}_{e^+e^-\to f}$
for these final states at $\sqrt{s}=$ 3.773 and 3.650 GeV, respectively.
They are also shown in the last columns of Tabs. \ref{tab:crs3773} and
\ref{tab:crs3650}.

\begin{table*}[hbtp]
\begin{center}
\caption{
The observed cross sections for $e^+e^-\to$ exclusive light hadrons at $\sqrt{s}=$
3.773 GeV, where
$N^{\rm obs}$ is the number of events observed from the $\psi(3770)$
resonance data, $N^{\rm b}$ is the total number of background events,
$N^{\rm net}$ is the number of the signal events,
$N^{\rm up}$ is the upper limit on the number of the signal events,
$\epsilon$ is the detection efficiency,
$\Delta_{\rm sys}$ is the relative systematic error in the measurement of
the observed cross section,
$\sigma$ is the observed cross section and
$\sigma^{\rm up}$ is the upper limit on the observed cross section set at
90\% C.L.}
\begin{tabular}{|l|c|c|c|c|c|c|} \hline
\multicolumn{1}{|c|}{$e^+e^-\to$} &$N^{\rm obs}$&$N^{\rm b}$
&$N^{\rm net}$ (or $N^{\rm up}$)
&$\epsilon$[\%]&$\Delta_{\rm sys}$
&$\sigma$ (or $\sigma^{\rm up}$) [pb] \\ \hline
$\kk2(\pp)$      &$ 92.7\pm 9.7$&$ 3.2\pm0.5$&$ 89.5\pm 9.7$&$ 3.08\pm0.06$&0.141&$168.0\pm18.2\pm23.7$\\
$2(\kk)\pp$      &$  5.0\pm 2.3$&$ 0.3\pm0.1$&$  4.7\pm 2.3$&$ 2.29\pm0.05$&0.143&$ 11.9\pm 5.8\pm 1.7$\\ 
$p\bar p2(\pp)$  &$ 24.0\pm 4.9$&$ 0.9\pm0.2$&$ 23.1\pm 4.9$&$ 5.69\pm0.08$&0.150&$ 23.5\pm 5.0\pm 3.5$\\ 
$4(\pp)$         &$ 50.0\pm 7.1$&$ 1.9\pm0.3$&$ 48.1\pm 7.1$&$ 2.11\pm0.05$&0.179&$131.8\pm19.5\pm23.6$\\ 
$\kk2(\pp)\pi^0$ &$ 21.3\pm 5.0$&$ 3.1\pm0.6$&$ 18.2\pm 5.0$&$ 0.46\pm0.02$&0.160&$231.5\pm63.6\pm37.0$\\ 
$4(\pp)\pi^0$    &6             &0           &$<11.47$      &$ 0.40\pm0.02$&0.189&$<206.9$             \\ 
$\rho^0\pp$      &$330.1\pm37.4$&$10.4\pm1.9$&$319.7\pm37.4$&$16.52\pm0.26$&0.117&$111.9\pm13.1\pm13.1$\\ 
$\rho^0\kk$      &$ 50.0\pm14.4$&$ 7.1\pm1.5$&$ 42.9\pm14.4$&$ 7.26\pm0.12$&0.128&$ 34.2\pm11.5\pm 4.4$\\ 
$\rho^0 p\bar p$ &$ 41.6\pm10.1$&$ 0.3\pm0.1$&$ 41.3\pm10.1$&$18.25\pm0.14$&0.141&$ 13.1\pm 3.2\pm 1.8$\\ 
$K^{*0}\kpi+c.c.$&$ 84.3\pm13.6$&$ 1.0\pm0.2$&$ 83.3\pm13.6$&$ 7.63\pm0.12$&0.110&$ 94.7\pm15.5\pm10.4$\\ 
$\ll$            &0             &0           &$<2.44$       &$15.77\pm0.13$&0.116&$<2.5$               \\
$\ll\pp$         &3             &0           &$<7.42$       &$ 4.63\pm0.07$&0.150&$<26.7$              \\ \hline
\end{tabular}
\label{tab:crs3773}
\end{center}
\end{table*}

\subsection{Upper limits on the observed cross sections and the
branching fractions for $\psi(3770)\to f$}
With the measured observed cross sections
$\sigma^{\rm 3.773\hspace{0.05cm}GeV}_{e^+e^- \to f}$ and
$\sigma^{\rm 3.650 \hspace{0.05cm}GeV}_{e^+e^- \to f}$
for $e^+e^- \to f$ at $\sqrt s=$ 3.773 and 3.650 GeV,
as listed in Tabs. \ref{tab:crs3773} and \ref{tab:crs3650},
we determine the observed cross section for $\psi(3770)\to f$ by
\begin{equation}
\sigma_{\psi(3770)\to f}=
\sigma^{\rm 3.773\hspace{0.05cm}GeV}_{e^+e^-\to f} -
f_{\rm co}\times \sigma^{\rm 3.650\hspace{0.05cm}GeV}_{e^+e^- \to f},
\label{eq:obscrs}
\end{equation}
where $f_{\rm co}$ is the coefficient due to the 1/s dependence of
the cross section. Here, we ignore the possible interference effects between
the continuum and resonance amplitudes, and neglect the difference of the
vacuum polarization corrections at the two energy points.
These yield $\sigma_{\psi(3770)\to f}$ for each mode, as shown in
the second column of Tab. \ref{tab:up_psipp},
where the first error is the statistical, the second is
the independent systematic arising from the uncertainties in the
Monte Carlo statistics, in fitting to the mass spectrum and in the
background subtraction, and the third is the common systematic error
arising from the other uncertainties as discussed in the subsection B.
For the final states with a few signal events observed from the continuum
data, such as $2(\kk)\pp$ and $4(\pp)\pi^0$, etc.,
we neglect their contributions from the continuum production in the
determination of $\sigma_{\psi(3770)\to f}$.

The upper limits on the observed cross sections,
$\sigma^{\rm up}_{\psi(3770) \to f}$, for $\psi(3770)$ decay to
$4(\pp)\pi^0$, $\ll$ and $\ll\pp$ are directly set based on the upper
limits on their observed cross sections at 3.773 GeV. While, 
for the other final states, the $\sigma^{\rm up}_{\psi(3770) \to f}$
is set through shifting the cross section by
1.64$\sigma$, where $\sigma$ is the total error of the
measured cross section. For the $\rho^0\kk$ and $\rho^0 p\bar p$ final
states, the central values of $\sigma_{\psi(3770)\to f}$ are less
than zero. In this case, we treat them as zero and then set their upper
limits. The results on $\sigma^{\rm up}_{\psi(3770) \to f}$ are summarized in
the third column of Tab. \ref{tab:up_psipp}.

The upper limit on the branching fraction for $\psi(3770)\to f$,
${\mathcal B}^{\rm up}_{\psi(3770)\to f}$, is set by dividing
its upper limit on the observed cross section
$\sigma^{\rm up}_{\psi(3770) \to f}$ by the observed cross section
$\sigma^{\rm obs}_{\psi(3770)}$ for the $\psi(3770)$ production at 3.773
GeV and a factor $(1-\Delta \sigma^{\rm obs}_{\psi(3770)})$,
where $\Delta \sigma^{\rm obs}_{\psi(3770)}$ is the total
relative error of the $\sigma^{\rm obs}_{\psi(3770)}$.
Here, $\sigma^{\rm obs}_{\psi(3770)}=(7.15\pm0.27\pm0.27)$ nb
\cite{crshads}, is obtained by weighting the two measurements
\cite{brdd2,rval} from BES Collaboration.
The results on ${\mathcal B}^{\rm up}_{\psi(3770)\to f}$ are shown
in the last column of Tab. \ref{tab:up_psipp}.
The upper limits on the observed cross sections and the branching fractions for
$\psi(3770)\to \ll$ and $\psi(3770)\to \ll\pp$ are consistent with the
measurements \cite{huang} from CLEO Collaboration.

\begin{table*}[hbtp]
\begin{center}
\caption{
The observed cross sections for $e^+e^-\to$ exclusive light hadrons at $\sqrt{s}=$
3.650 GeV, where
$N^{\rm obs}$ is the number of events observed from the continuum data,
and the definitions of the other symbols are the same as those in Tab.
\ref{tab:crs3773}.}
\begin{tabular}{|l|c|c|c|c|c|c|} \hline
\multicolumn{1}{|c|}{$e^+e^-\to$} &$N^{\rm obs}$&$N^{\rm b}$
&$N^{\rm net}$ (or $N^{\rm up}$)
&$\epsilon$[\%]&$\Delta_{\rm sys}$
&$\sigma$ (or $\sigma^{\rm up}$) [pb] \\ \hline
$\kk2(\pp)$      &$ 30.0\pm 5.5$&$0.1\pm0.0$&$ 29.9\pm 5.5$&$ 2.79\pm0.05$&0.141&$164.9\pm 30.3\pm23.2$\\
$2(\kk)\pp$      &2             &0          &$<5.91$       &$ 2.16\pm0.05$&0.142&$ <49.1$              \\
$p\bar p2(\pp)$  &$  8.0\pm 2.9$&$0.1\pm0.0$&$  7.9\pm 2.9$&$ 5.33\pm0.07$&0.150&$ 22.8\pm  8.4\pm 3.4$\\
$4(\pp)$         &$ 10.0\pm 3.2$&$0.0\pm0.0$&$ 10.0\pm 3.2$&$ 2.02\pm0.04$&0.179&$ 76.2\pm 24.4\pm13.6$\\
$\kk2(\pp)\pi^0$ &$  5.4\pm 2.5$&$0.0\pm0.0$&$  5.4\pm 2.5$&$ 0.56\pm0.02$&0.150&$150.2\pm 69.5\pm22.5$\\
$4(\pp)\pi^0$    &0             &0          &$<2.44$       &$ 0.39\pm0.01$&0.184&$<119.4$              \\
$\rho^0\pp$      &$133.5\pm24.4$&$3.3\pm0.8$&$130.2\pm24.4$&$17.63\pm0.28$&0.115&$113.6\pm 21.3\pm13.1$\\
$\rho^0\kk$      &$ 30.4\pm 8.8$&$2.0\pm0.6$&$ 28.4\pm 8.8$&$ 7.58\pm0.12$&0.110&$ 57.6\pm 17.9\pm 6.3$\\
$\rho^0 p\bar p$ &$ 22.2\pm 7.8$&$0.0\pm0.0$&$ 22.2\pm 7.8$&$19.32\pm0.14$&0.157&$ 17.7\pm  6.2\pm 2.8$\\
$K^{*0}\kpi+c.c.$&$ 29.5\pm 9.0$&$0.2\pm0.1$&$ 29.3\pm 9.0$&$ 7.91\pm0.12$&0.169&$ 85.5\pm 26.3\pm14.4$\\
$\ll$            &0             &0           &$<2.44$     &$16.97\pm0.13$&0.116&$<6.1$                \\
$\ll\pp$         &1             &0           &$<4.36$     &$ 4.51\pm0.07$&0.150&$<42.9$               \\
\hline
\end{tabular}
\label{tab:crs3650}
\end{center}
\end{table*}

\newcommand{\rb}[1]{\raisebox{1.5ex}[0pt]{#1}}
\begin{table*}[htbp]
\begin{center}
\caption{
The upper limits on the observed cross section $\sigma^{\rm up}_{\psi(3770)\to
f}$ and the branching fraction ${\mathcal B}^{\rm up}_{\psi(3770)\to f}$
for $\psi(3770)\to f$ are set at 90\% C.L. The $\sigma_{\psi(3770)\to f}$
in the second column is calculated with Eq. (\ref{eq:obscrs}), where the first
error is the statistical, the second is the independent systematic, and the
third is the common systematic error. Here, the upper $^t$ denotes that we
treat the upper limit on the observed cross section for $e^+e^-\to f$ at 3.773 GeV
as $\sigma^{\rm up}_{\psi(3770)\to f}$, the upper $^n$ denotes that we neglect the
contribution from the continuum production, and the upper $^z$ denotes that
we treat
the central value of $\sigma_{\psi(3770)\to f}$ as zero if it is less
than zero.
}
\begin{tabular}{|l|c|c|c|} \hline
   & $\sigma_{\psi(3770)\to f}$
   & $\sigma^{\rm up}_{\psi(3770)\to f}$
   & ${\mathcal B}^{\rm up}_{\psi(3770)\to f}$ \\
\multicolumn{1}{|c|}{\rb{Decay Mode}} & [pb] & [pb]    & [$\times 10^{-3}$] \\ \hline
$\kk2(\pp)$      &$ 13.7\pm 33.7\pm 4.4\pm 1.9$    &$< 69.5$&$<10.3$\\
$2(\kk)\pp$      &$ 11.9\pm5.8\pm0.4\pm1.7^{n}$    &$< 21.8$&$< 3.2$\\
$p\bar p2(\pp)$  &$  2.2\pm  9.3\pm 0.5\pm 0.3$    &$< 17.5$&$< 2.6$\\
$4(\pp)$         &$ 60.5\pm 30.0\pm 3.6\pm10.7$    &$<113.1$&$<16.7$\\
$\kk2(\pp)\pi^0$ &$ 90.9\pm 91.0\pm16.1\pm13.2$    &$<244.0$&$<36.0$\\
$4(\pp)\pi^0$    &$<206.9^{tn}$                    &$<206.9$&$<30.6$\\
$\rho^0\pp$      &$  5.6\pm 23.9\pm 7.6\pm 0.6$    &$< 46.6$&$< 6.9$\\
$\rho^0\kk$      &$-19.7\pm 20.3\pm 3.1\pm 2.1^z$  &$< 33.9$&$< 5.0$\\
$\rho^0 p\bar p$ &$ -3.5\pm  6.6\pm 2.1\pm 0.4^z$  &$< 11.4$&$< 1.7$\\
$K^{*0}\kpi+c.c.$&$ 14.7\pm 29.1\pm11.1\pm 1.5$    &$< 65.8$&$< 9.7$\\
$\ll$            &$<2.5^{tn}$                      &$<2.5$  &$< 0.4$\\
$\ll\pp$         &$<26.7^{tn}$                     &$<26.7$ &$< 3.9$\\
\hline
\end{tabular}
\label{tab:up_psipp}
\end{center}
\end{table*}

\section{Summary}
In summary, by analyzing the data sets taken at $\sqrt{s}=$ 3.773 and
3.650 GeV with the BESII detector at the BEPC collider, we have measured
the observed cross sections for 12 exclusive light hadron final states
produced in $e^+e^-$ annihilation at the two energy points. With the
measured observed cross sections at the two energy points, we set the upper
limits on the observed cross sections and the branching fractions for
$\psi(3770)$ decay to these final states at 90\% C.L. In the
measurements, we ignore the interference effects between the continuum and
resonance amplitudes due to not knowing the details about the two amplitudes.
In this case, even though we have not observed significant
difference between the observed cross sections for most light
hadron final states at the two energy points, we still can not conclude
that the $\psi(3770)$ does not decay into these final states.
To extract the branching fractions for $\psi(3770)\to$ exclusive light
hadrons, a better way is to analyze their energy-dependent observed cross
sections at more energy points covering both $\psi(3770)$ and $\psi(3686)$
\cite{crsscan}.
However, these observed cross sections reported in this
Letter and those reported in Ref. \cite{crshads}, would be valuable to
get better comprehensions for both the mechanism of the continuum
light hadron production and the discrepancy between the observed cross
sections for $D\bar D$ and $\psi(3770)$ production.

\section{acknowledgments}
The BES collaboration thanks the staff of BEPC for their hard
efforts. This work is supported in part by the National Natural
Science Foundation of China under contracts Nos. 10491300,
10225524, 10225525, 10425523, the Chinese Academy of Sciences
under contract No. KJ 95T-03, the 100 Talents Program of CAS under
Contract Nos. U-11, U-24, U-25, the Knowledge Innovation Project
of CAS under Contract Nos. U-602, U-34 (IHEP), the National
Natural Science Foundation of China under Contract  No. 10225522
(Tsinghua University).

\end{document}